\documentclass[10pt,journal,compsoc]{IEEEtran}

\usepackage{color}
\usepackage{times}
\usepackage{graphicx}
\usepackage{graphicx,epstopdf}
\usepackage{dcolumn}
\usepackage{bm}
\usepackage{mathrsfs}
\usepackage{soul}
\usepackage{amsfonts}
\usepackage{float}

\usepackage{mathtools}
\usepackage{blkarray, bigstrut}
\usepackage{hyperref}

\usepackage{mathtools}

\ifCLASSOPTIONcompsoc
  \usepackage[nocompress]{cite}
\else
  \usepackage{cite}
\fi
\ifCLASSINFOpdf
\else
\fi
\begin{document}
%
\title{Distance dependent competitive interactions in a frustrated network of mobile agents}
%
%
%
%

\author{Sayantan~Nag~Chowdhury,
        Soumen~Majhi,
        and~Dibakar~Ghosh
\IEEEcompsocitemizethanks{\IEEEcompsocthanksitem S. Nag Chowdhury, S. Majhi and D. Ghosh are with the Physics and Applied Mathematics Unit, Indian Statistical Institute, 203 B. T. Road, Kolkata 700108, India.\protect\\
E-mail: dibakar@isical.ac.in (D. Ghosh)
}}
\IEEEtitleabstractindextext{%
\begin{abstract}
Diverse collective dynamics emerge in dynamical systems interacting on top of complex network architectures. Along this line of research, temporal network has come out to be one of the most promising network platforms to investigate. Especially, such network with spatially moving agents has been established to be capable of modelling a number of practical instances. In this paper, we examine the dynamical outcomes of moving agents interacting based upon their physical proximity. For this, we particularly emphasize on the impact of competing interactions among the agents depending on their physical distance. We specifically assume attractive coupling between agents which are staying apart from each other, whereas we adopt repulsive interaction for agents that are sufficiently close in space. With this set-up, we consider two types of coupling configurations, symmetry-breaking and symmetry-preserving couplings. We encounter variants of collective dynamics ranging from synchronization, inhomogeneous small oscillation to cluster state and extreme events while changing the attractive and repulsive coupling strengths. We have been able to map all these dynamical behaviors in the coupling parameter space. Complete synchronization being the most desired state in absence of repulsive coupling, we present an analytical study for this scenario that agrees well with the numerical results. 
\end{abstract}

\begin{IEEEkeywords}
Time varying networks, Mobile agents, Extreme events, Synchronization, Inhomogeneous small oscillation.
\end{IEEEkeywords}}

\maketitle

\IEEEdisplaynontitleabstractindextext

%
\IEEEpeerreviewmaketitle

\IEEEraisesectionheading{\section{Introduction}	\label{intro}}

\IEEEPARstart{C}{omplex} network \cite{newman_10,barabasi2016network} is the unifying paradigm behind many scientific disciplines ranging from physics, mathematics to computer science and biology. Besides the study related to local and global statistical properties of complex networks \cite{albert2002statistical}, researchers are equally interested in paying attention to the dynamics of their interacting units \cite{boccaletti2006complex}. Among those collective dynamics, a widely studied phenomenon is that of synchronization \cite{arenas2008synchronization,rakshit2020invariance,wang2002synchronization,vaidyanathan2014adaptive,rakshit2020intralayer,vaidyanathan2015analysis,wu1996conjecture,rakshit2018synchronization} of coupled oscillators arranged over the constituents of complex networks. 
Although, in recent times, there have been attempts in examining synchronizability of time-varying networks with different underlying structural arrangements, in most of these studies of synchronization, the nodes of the complex networks are spatially static. But, in real world networks, the individuals (nodes)  move around and exchange information within close proximity. Such nodes in the contextual literature are well known as \textit{mobile agents} \cite{kerr2002local,reichenbach2007mobility,song2010limits,buscarino2016interaction,stilwell2006sufficient,nag2020cooperation,frasca2008synchronization,fujiwara2011synchronization,uriu2013dynamics} 
referring to those nodes whose movement highly influence the systems' collective dynamics. But, in general, the states of those oscillators situated on top of those agents do not affect the agents' mobility. The studies on the process of synchronization in mobile agents \cite{majhi2017synchronization,uriu2010random,chowdhury2019convergence,porfiri2006random,chowdhury2019synchronization1,kim2013emergence,majhi2019emergence} have, so far,  primarily concentrated on those particular cases when the mobile agents interact with each other only when they are closed enough in space. But, in all of these studies, the network architecture, i.e.,\  how the oscillators are connected with each other, depends only on the attractive interaction. So, a basic query arises that how is the flow of information gets affected when different types of interactions are present in the movement. For instance, an attempt to answer this question can be made while incorporating repulsive coupling with the attractive one in such a networked system. Such co-existing dynamical interactions can be observed in a variety of systems and have been studied previously in different network set-ups yielding diverse collective states  \cite{bera2016emergence,mishra2015chimeralike,wang2011synchronous,hens2014diverse,tian2009interlinking}.

\par Synergy of attractive-repulsive coupling can lead to several interesting dynamical behaviors. One of such example is the emergence of solitary states in multiplex networks \cite{majhi2019solitary} possessing positive inter-layer and negative intra-layer interactions. Earlier, Maistrenko et al.\ \cite{maistrenko2014solitary} found a variety of stationary states in networks of globally coupled identical oscillators with attractive and repulsive interactions. Sathiyadevi et al.\ \cite{sathiyadevi2018distinct,sathiyadevi2018stable} found diverse chimera states  due to the introduction of nonlocal repulsive coupling together with an attractive coupling in a network of coupled oscillators. Also, interplay between “conformists” and “contrarians” leads to traveling wave along with the stationary states \cite{hong2011kuramoto}. The effect of competitive interactions on the synchronization manifold among globally coupled identical Van der Pol oscillators  through their velocities is rigorously addressed in ref.\ \cite{vaz2011synchronisation}. Even, the presence of a tiny fraction of repulsive couplings is found to enhance the synchronization of nonidentical dynamical units that are attractively coupled in a small-world network \cite{leyva2006sparse}. Origination of dragon-king-like extreme events as a result of the coexistence of excitatory and inhibitory chemical synaptic couplings is reported in the ref. \cite{mishra2018dragon}. Besides emergent dynamical patterns, co-presence of attractive and repulsive couplings
carry high significance to several biological \cite{inoue2010dynamics}, physical \cite{ottino2018volcano}, ecological \cite{bacelar2014exploring} and even social scenarios \cite{szell2010multirelational}. 

\par Recently, Nag Chowdhury et al.\ \cite{chowdhury2020effect} proposed an universal $0-\pi$ rule to identify the bipartiteness of networks from the anti-phase synchronization states. This article describes such a system of attractively coupled oscillators with negative links as \textit{frustrated networks} \cite{gade2013frustration}. 
Such frustrated systems under the influence of attractive-repulsive co-existing interactions may induce unanticipated phenomenon, like \textit{extreme events} \cite{lucarini2016extremes}.  Most of the existing articles \cite{ray2019intermittent,ansmann2013extreme,mishra2018dragon,ray2020extreme,ray2020understanding} on extreme events in dynamical systems and networks define such an event as rare event. Mathematically, a rare event is defined on a probability space $\chi$ equipped with a $\sigma$-algebra $\mathbb{B}$ of events such that $P(A)$ is small, where $P$ is the probability
measure designed to quantify the probability of the occurrence of any event $A \in \mathbb{B}$. But, extreme events can be rare  events \cite{cousins2015unsteady,guth2019machine,dematteis2018rogue}, or they can often be frequent in time and space \cite{grooms2014stochastic,majda2014conceptual,cai2001dispersive,majda2015intermittency}. Examples of such intermittent frequent extreme events include population size of Paris \cite{sornette2009dragon}, the bubbling of share market \cite{feigenbaum2001statistical} before a crash and many more. 

\par Along with its small occurrences, extreme event has another threat in the form of large impact. Financial and commodity market crashes, tsunamis, hurricanes, floods, epidemic disease spread,
global warming-related changes in climate and weather, warfare and related forms of violent conflict,
asteroid impacts, solar flares, acts of terrorism, industrial accidents, 8.0+ Richter magnitude earthquake are few examples of such large impact events. 
In dynamical systems, researchers are recently starting to measure the impact of such events in terms of the amount of variation from the central tendency of that observable \cite{ray2019intermittent,mishra2018dragon,ray2020extreme,bonatto2011deterministic,dysthe2008oceanic,slunyaev2009rogue}. So, if $\psi$ is the observable defined on the state space $U$ to $\mathbb{R}$, then the extreme events are members of the set $\{u \in U : \psi(u) > H_S\}$, where $H_S=m+d \sigma$ with $d$ is the non-zero integer. Here, $m$ and $\sigma$ are respectively the mean value and the standard deviation of all the peak values in a time series of $u$. But, if $H_S$ is too large, there will be very few values to model the tail of the distribution correctly as the variance is likely to be large due to only very extreme observations remaining. On the other
hand, a low threshold value of $H_S$ will include too many values giving a high bias. Recently, $m+8 \sigma$ is found to appropriate indicating extreme event threshold for Weibull distribution \cite{chowdhury2019synchronization}.

\par Inspired by these observations, we consider a frustrated network of limit cycle oscillators, where each node of the network is a mobile agent. Earlier all studies \cite{buscarino2006dynamical,majhi2017synchronization,uriu2010random,porfiri2006random,kim2013emergence,majhi2019emergence,chowdhury2019synchronization} related to mobile agents, considered only limited local interactions. Instead, we consider here a global network of mobile agents with co-existing switching interactions. Depending solely on the relative distance between the agents, attractive or repulsive interaction is activated. Specifically, we choose attractive interaction between those agents which are staying apart from each other, while we consider repulsive coupling for the agents that are sufficiently close. We consider two types of coupling schemes, namely symmetry-breaking and symmetry-preserving couplings. Under this set up, we explore the effect of coupling strengths which reveal quite interesting phenomena like \textit{cluster synchronization} \cite{pecora2014cluster}, \textit{complete synchronization}, \textit{inhomogeneous small oscillation}, \textit{extreme events} etc. We also analytically derive the critical coupling strength for achieving complete synchronization using time-average Laplacian matrix in absence of repulsive interaction and numerically verify the results. 




\section{Mathematical framework} \label{Mathematical framework}

\begin{figure}[ht]
	\centerline{
		\includegraphics[scale=0.6]{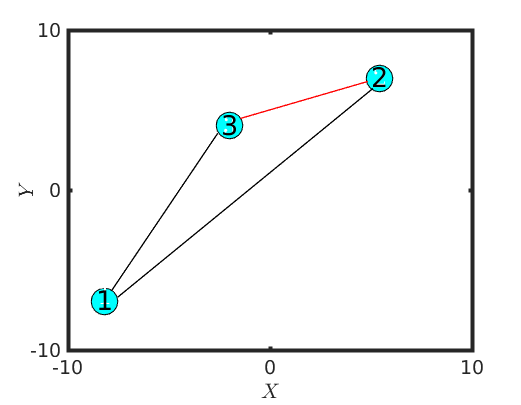}}
	\caption{Schematic diagram at a particular time: We consider $N=3$ mobile agents in the two-dimensional XY-plane, where $g=10$ and $\alpha=10$. The red and black lines represent repulsive and  attractive links, respectively.}
	\label{picture1}
\end{figure}

We consider $N$ mobile agents which are moving independently on a two-dimensional ($2$D) physical plane $S=[-g,g] \times [-g,g]$. Initially, the agents are randomly distributed on $S$. The $i$-th mobile agent can move in any direction with velocity $\big(v\cos\theta_{i}(t), v\sin\theta_{i}(t)\textbf{})$ (any kind of collision among the agents are not allowed), where $v$ is the uniform moving velocity and $\theta_{i}$ is drawn randomly from $[0, 2\pi]$. Higher values of velocity $v$ implies that the agents move the whole physical space and which increases the possibilities of interactions.  Thus, if $\big(p_{i}(t),q_{i}(t)\big)$ is the position of the $i$-th agent at time $t$, then motion updating process will be maintained by the following relation,

\begin{equation}\label{1}
\begin{split}
p_{i}(t+1)=p_{i}(t)+v\cos\big(\theta_{i}(t)\big),\\
q_{i}(t+1)=q_{i}(t)+v\sin\big(\theta_{i}(t)\big).
\end{split}
\end{equation}
\begin{figure*}[ht]
	\centerline{
		\includegraphics[scale=0.6]{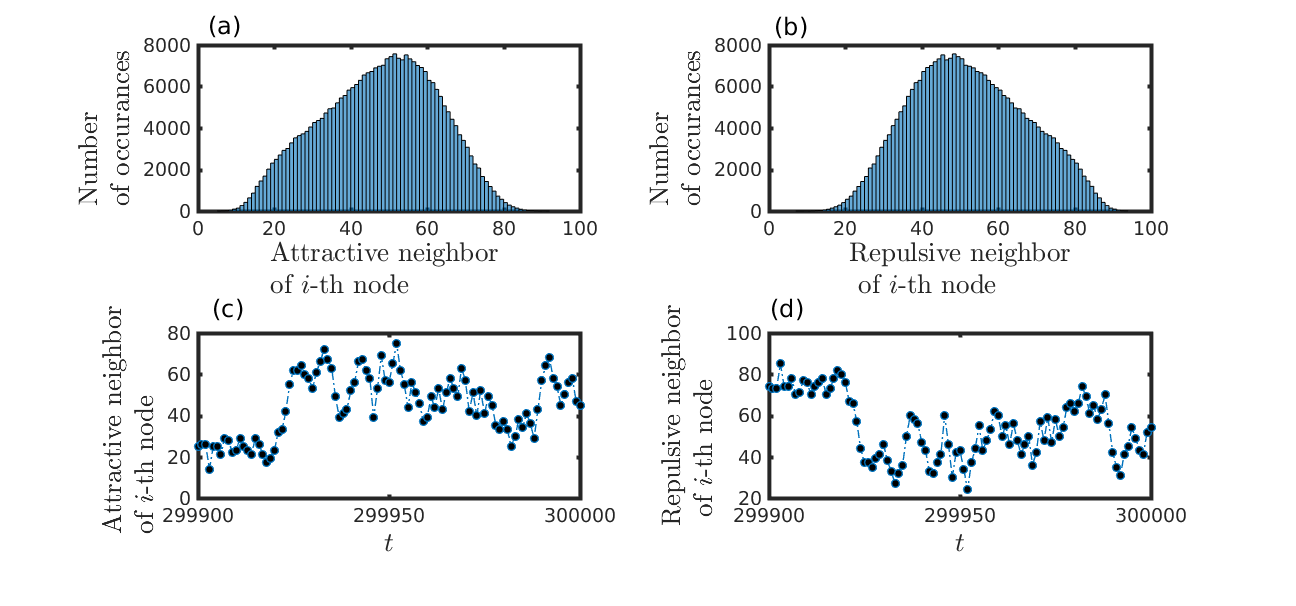}}
	\caption{Variation of degree distribution of a randomly chosen $i$-th agent out of $N=100$ mobile agents: (a) attractive neighbor and (b) repulsive neighbor.  Each $i$-th agent is connected with remaining $(N-1)$ agents whenever $\alpha \in [0, 2\sqrt{2}g]$. Note that the numbers of attractive and repulsive neighbors both are widely spread. The simulation is carried out for $t=3 \times 10^5$. (c, d) Variation of degree of both matrices with respect to time $t$: Just like the upper panel, we find that the neighbors of any agent are rapidly changing depending on the agent's random movement. Here, we take $g=10$ and $\alpha=10$, and the uniform moving velocity $v=2.0$.}
	\label{picture2}
\end{figure*}
If at some time $t$, $p_{i}(t),q_{i}(t)$ exceed $|g|$, then we re-generate a new $\theta_{i}(t)$ in $[0, 2 \pi]$ such that $-g \leq p_{i}(t),q_{i}(t) \leq g$. We define the distance between any two agents by the standard Euclidean metric as

\begin{equation}\label{2}
d_{ij}(t)=\sqrt{\Big(p_{i}(t)-p_{j}(t)\Big)^2+ \Big(q_{i}(t)-q_{j}(t)\Big)^2}.
\end{equation}
Next, we write the the dynamical equations characterizing the time-evolution of each agent ($i=1,2,\cdots,N$) in the networked system as follows,

\begin{equation}\label{4}
\dot{\mathbf{x}}_{i}=F(\mathbf{x_{i}})+\sum_{j=1}^{N} \{k_{A} B_{ij}+k_{R} C_{ij} \}H(\mathbf{x_{i}},\mathbf{x_{j}}),
\end{equation}
where $\mathbf{x}_i \in \mathbb{R}^n$ is the state variable and  $F(\mathbf{x_{i}})$ is the corresponding vector field of the $i$-th agent.  Here $H(\mathbf{x_{i}},\mathbf{x_{j}})$ is chosen as the diffusive type coupling function, $k_A > 0$ and $k_R < 0$ respectively correspond to the attractive and repulsive coupling strengths. 
Moreover, $B$ and $C$ are the distance dependent adjacency matrices associated to the attractive and repulsive interactions such that 

\[B_{ij}=\begin{cases}
1, & \text{if} \hspace{0.1cm}  d_{ij}(t) > \alpha \\
0, & \text{if} \hspace{0.1cm} d_{ij}(t) \leq \alpha,
\end{cases}
\]
and 
\[C_{ij}=\begin{cases}
0, & \text{if} \hspace{0.1cm}  d_{ij}(t) > \alpha \\
1, & \text{if} \hspace{0.1cm} d_{ij}(t) \leq \alpha,
\end{cases}
\]
where $\alpha$ is the parameter lying within $[0, 2\sqrt{2}g]$. Here $2\sqrt{2}g$ is the maximum possible distance between any two agents lying in the $2$-dimensional plane $[-g,g] \times [-g,g]$. At a particular instance, the agents whose Euclidean distance is less than $\alpha$, repel to each other and repulsive interactions arise between them. Attractive coupling occurs between the agents when the distance between them  is greater than $\alpha$.

\par To illustrate this, we draw a schematic diagram with $N=3$ mobile agents in the $2$-dimensional plane $[-g,g] \times [-g,g]$ in Fig.\ (\ref{picture1}). At a particular time $t$, we plot the position of these three agents in Fig.\ (\ref{picture1}) where $g=10$ and $\alpha=10$ are considered. For these choices of $\alpha$ and $g$, the adjacency matrices corresponding to attractive and repulsive interactions at that specific time instant look like 

\begin{center}
	
$ B=\left(\begin{array}{ccccc}
0 & 1 & 1\\
1 & 0 & 0\\
1 & 0 & 0
\end{array}\right)$  and
\nolinebreak
$ C=\left(\begin{array}{ccccc}
0 & 0 & 0\\
0 & 0 & 1\\
0 & 1 & 0
\end{array}\right)$\\
\end{center}
\par Now, we consider fix physical plane with $g=10$ and fix the parameter $\alpha=10.$ Each agents moving within the physical plane $[-g,g] \times [-g,g]$ using the rule given in (\ref{1}) with fixed velocity $v=2.0$. Now we see the variation of degree distributions of the time-varying network. For this, we calculate the number of attractive and repulsive neighbors depending on the critical distance $\alpha$.  In Fig.\ \ref{picture2}, the degree distributions of moving agents for attractive and repulsive interactions are shown.  We randomly select an agent which is moving with uniform velocity $v=2.0$. We then find that the number of attractive neighbors are varying within the interval $[5,92]$ (cf. Fig.\ \ref{picture2}(a)) and the number of repulsive neighbors of that agent is varying within $[7,94]$ (cf. Fig.\ \ref{picture2}(a)). At any particular time-step, the sum of the number of attractive neighbors and the number of repulsive neighbors is $N-1$ for any $i$-th agent. This scenario is well expressed by the lower panel of Figs.\ \ref{picture2}(c, d).

\par Now, on top of each mobile agent, a two-dimensional Stuart-Landau (SL) oscillator is placed where the state dynamics of each limit cycle oscillator is represented by

\begin{equation}\label{3}
\begin{split}
F(\mathbf{x_{i}})=\left(
\begin{array}{c}
\left[1-\left({x_i}^2+{y_i}^2\right)\right]x_i-\omega_i y_i\\\\
\left[1-\left({x_i}^2+{y_i}^2\right)\right]y_i+\omega_i x_i\\
\end{array}
\right), 
\end{split}
\end{equation}
where $\omega_{i}=\omega=3.0$ is the identical intrinsic frequency for each $i=1, 2, \cdots, N$. 


\par We consider two types of diffusive interactions, namely symmetry-breaking coupling $H(\mathbf{x_{i}},\mathbf{x_{j}})=({x_{j}}-{x_{i}},0)^T$ and symmetry-preserving coupling $H(\mathbf{x_{i}},\mathbf{x_{j}})=({x_{j}}-{x_{i}}, {y_{j}}-{y_{i}})^T$. The interesting and important part of this distance dependent coupling functions is for $\alpha = 0,$ the Eq. \eqref{4} becomes globally coupled network with attractive coupling only. In this case, movement of the agents does not effect the attractive adjacency matrix $B$ and the network becomes static.  But, for $\alpha>0,$ depending upon the values of attractive and repulsive coupling strengths $k_A$ and $k_R$, two different cases can be implemented:
\begin{enumerate}
	\item $k_A \ne 0$ and $k_R=0:$ In this case,  the mobile agents will interact attractively when the relative distance $d_{ij}$ is greater than $\alpha$ and remain disconnected with each other when they are closed to each other and relative distance $d_{ij}$ is less than equal to $\alpha$. There is no repulsion among the agents. Here, synchronization is the most desired state for a suitable value of $k_A.$
	\item $k_A \ne 0$ and $k_R\ne 0:$ In this case, interplay between attractive and repulsive interactions is considered. During the spatial movement of the agents, they experience both types of interaction among them. Here, different states may arise, like inhomogeneous small oscillation, synchronization, extreme event and other intermittent states.  
\end{enumerate} 

\par In the following section, we will explore the dynamics of the time-varying network \eqref{4} with two types of diffusive couplings and under above mentioned distinct coupling conditions. Our main emphasis will be to identify the parameter space by varying the two coupling strengths $k_A$ and $k_R$ with fixed values of other network parameters $g= \alpha=10.0$ and $v=2.0$ related to the movement of the agents.

\section{Results} \label{Results}
 For numerical simulations, we integrate Eq.\ (\ref{4}) using fifth order Runge-Kutta-Fehlberg method with a integration time step $\delta t=0.01$. At each integrating time step, $\theta_{i}(t)$ and thus $v_{i}(t)$ is updated according to rule given in (\ref{1}) for each $i$-th mobile agent ($i=1,2,\cdots,N$). Hence, both the matrices $B$ and $C$ change at each integrating time step leading to fast switching approximation \cite{stilwell2006synchronization}. All simulations are done for fixed initial conditions 
\begin{equation}\label{5}
\begin{split}
x_i(0)=(-1)^i \frac{i}{N},\\
y_i(0)=(-1)^i \frac{i}{N}, \\
\end{split}
\end{equation}
unless stated otherwise \footnote{ Inhomogeneous small oscillation depends crucially on the initial conditions, and the basin of attraction of this state is small. If the initial conditions are not chosen suitably from the basin of attraction, then the trajectories may converge to a different periodic attractor. However, other observed dynamical states can be reached from any random initial conditions starting from $[-1,1] \times [-1,1]$.}. In this Sec.\ (\ref{Results}), the symmetry-breaking coupling $H(\mathbf{x_{i}},\mathbf{x_{j}})=({x_{j}}-{x_{i}},0)^T$ is considered.



\subsection{Absence of repulsive interaction}

For the chosen values of the parameters as mentioned above, we find different dynamical behaviors of the network system \eqref{4}. We first look at the collective dynamics of the network whenever there is no repulsion among the agents and only attractive coupling is activated, i.e., when $k_A \neq 0$ and $k_R=0$. In this scenario, the agents with relative distance $d_{ij}(t) \leq \alpha=10$ remain disconnected with each other and agents will interact with each other if their distance is greater than $\alpha$.

\par We next intend to analytically derive the critical interaction strength $k_{critical}$ for which synchrony appears, based on the approach of constructing the time-average Laplacian matrix $ \overline{G} = [\overline{g_{ij}}] $. 
For the sake of simplicity, let us start with $N=2$ mobile agents. Then, depending on their relative distance, two possible Laplacian matrices can be observed. 

\begin{enumerate}
	\item Whenever $d_{ij}>\alpha$, then the agents interact with one another so that the corresponding  Laplacian matrix becomes 
	$ G_A=
	\left( \begin{array}{ccccc}
	1 & -1\\
	-1 & 1 
	\end{array} \right)
	$.
	\item If $d_{ij} \leq \alpha$, then the Laplacian matrix due to absence of individuals' interaction takes the form\\\\
	$ G_0= \left( \begin{array}{ccccc}
	0 & 0\\
	0 & 0
	\end{array}\right)$.
	
\end{enumerate}

\par Since $G_0$ is a null matrix so in this case $\overline{G}$ reduces to $ \overline{G} = pG_A $, where $p$ is the probability of interaction between the two agents. Since, we are considering a planar space $S$ of area $4g^2$ sq.\ units, then $p=1-\frac{\text{Area of the interaction }}{\text{Area of $S$}}$ becomes $p=1-\frac{\pi \alpha^2}{4g^2}$.

\par We next calculate the average Laplacian matrix $\overline{G}$ for $N=3$ agents. For instance, depending on the spatial positions of these mobile agents for a fixed value of $\alpha$, eight possible configurations have been encountered for $N=3$. These possible cases are described below:
\begin{enumerate}
	\item  All the agents interact with each other and thus $ G_{A}=\left( \begin{array}{ccccc}
	2 & -1 & -1\\
	-1 & 2 & -1\\
	-1  & -1 & 2
	\end{array} \right)$.

	\item There are three cases in which two out of the three agents interact with each other and the third one remains isolated. The corresponding Laplacian matrices are \\\\
	$ G_{12}=\left( \begin{array}{ccccc}
	1 & -1 & 0\\
	-1 & 1 & 0\\
	0  & 0 & 0
	\end{array}\right)$, 	$ G_{13}=\left( \begin{array}{ccccc}
	1 & 0 & -1\\
	0 & 0 & 0\\
	-1  & 0 & 1
	\end{array}\right)$ and 	$ G_{23}=\left( \begin{array}{ccccc}
	0 & 0 & 0\\
	0 & 1 & -1\\
	0  & -1 & 1
	\end{array}\right)$, where $G_{ij}$ is the Laplacian matrix when $i$-th and $j$-th agents are interacting with each other, and the other third agent remains isolated. Also note that, $ G_{12} + G_{13} + G_{23} = G_{A} $.
	
	\item It is also possible that two of the agents lie within the relative distance $\alpha$ so that they are not interacting with each other, but the third agent stays far away and interacts with both of them. The associated Laplacians are as follows\\\\
	$ G_{1}=\left( \begin{array}{ccccc}
	2 & -1 & -1\\
	-1 & 1 & 0\\
	-1  & 0 & 1
	\end{array}\right)$, 
	$ G_{2}=\left( \begin{array}{ccccc}
	1 & -1 & 0\\
	-1 & 2 & -1\\
	0  & -1 & 1
	\end{array}\right)$ and
	$ G_{3}=\left( \begin{array}{ccccc}
	1 & 0 & -1\\
	0 & 1 & -1\\
	-1  & -1 & 2
	\end{array}\right)$, where $G_{k}$ is the Laplacian matrix when the $k$-th agent interacts with the other two agents, but those two agents do not interact with each other. Here note that $ G_{1} + G_{2} + G_{3} = 2G_{A} $.
	
	\item Majority of the agents, i.e., all three oscillators situated within the relative distance $\alpha$, hence no one can establish a connection with the others. So, Laplacian matrix can be indicated by \\\\
	$ G_0=\left( \begin{array}{ccccc}
	0 & 0 & 0\\
	0 & 0 & 0\\
	0 & 0 & 0
	\end{array}\right)$.
	
\end{enumerate}

Then the time-average matrix for $N=3$ becomes, $\overline{G} = p_0G_0+p_{12}G_{12}+p_{13}G_{13}+p_{23}G_{23}+p_{1}G_{1}+p_{2}G_{2}+p_{3}G_{3}+p_AG_A$. Here $p$'s stand for the probabilities associated to the respective network configurations for which we have $p_{1}=p_{2}=p_{3}=p_{a}$(say), and $p_{12}=p_{13}=p_{23}=p_{ab}$(say). Then $\overline{G} = (2p_{a}+p_{ab}+p_A)G_A$ and hence $ \overline{G} = pG_A$, where $p=2p_{a}+p_{ab}+p_A$ is the probability of interaction between any two agents. The scenarios corresponding to $N \ge 4$ can be similarly tackled, all of which yields the time-average matrix as $ \overline{G}=p G_A$, where $G_A$ is the Laplacian matrix of order $N\times N$ in the form,
\\\\
\noindent
$ G_A=\left(\begin{array}{cccccccc}
N-1 & -1 & -1 & \cdots & \cdots & \cdots &-1 \\
-1 & N-1 & -1 & \cdots & \cdots & \cdots &-1 \\
-1 & -1 & N-1 & \cdots & \cdots & \cdots & -1\\
\cdots & \cdots & \cdots & \cdots & \cdots. & \cdots & \cdots\\
\cdots & \cdots & \cdots & \cdots & \cdots & \cdots & \cdots\\
\cdots & \cdots & \cdots & \cdots & \cdots & \cdots & \cdots\\
-1 & -1 & -1 & \cdots & \cdots & \cdots & N-1
\end{array}\right)$,
so that $\overline{G}$ is basically a rescaled all-to-all global Laplacian matrix.

\par As per our described model, the links among the agents get rewired at each integrating time step, so the convergence of all the attractors to a single attractor \cite{Nag_Chowdhury_2020} can be assured if the time-average matrix supports synchronization \cite{stilwell2006sufficient}. As stated in the Ref.\ \cite{stilwell2006sufficient}, if there exists a constant $T$ such that $G(t)$ satisfies $\frac{1}{T} \int_{t}^{t+T} G(\tau) d\tau = \overline{G} $, the time-average of $G(t)$ and the system of coupled oscillators given by
\begin{equation} \label{eq:7}
\begin{array}{lcl}
\dot{\mathbf{x}}_i=F(\mathbf{x_i})-k \sum_{j=1}^{N} \overline{g_{ij}} H(\mathbf{x_j}),~~~~i=1,2,...,N
\end{array}
\end{equation}
with fixed interacting Laplacian matrix $ \overline{G} = [\overline{g_{ij}}] $ possesses a stable synchronization manifold, then there exists $ {\epsilon}^{*} > 0 $ such that  for all fixed $ \epsilon $ satisfying $ 0 < \epsilon < {\epsilon}^{*}$, the  coupled system according to the time-varying Laplacian $ G(\frac{t}{\epsilon}) $ defined by
\begin{equation} \label{eq:8}
\begin{array}{lcl}
\dot{\mathbf{x_i}}=F(\mathbf{x_i})-k \sum_{j=1}^{N} g_{ij}(\frac{t}{\epsilon}) H(\mathbf{x_j}),~~~~i=1,2,...,N
\end{array}
\end{equation}
also sustains a stable synchrony manifold under sufficiently fast switching sequence between the network configurations.

\par Then, the stability of the synchronized state can be investigated by the eigen values of $ \overline{G} $. For a time-independent coupling matrix, a necessary condition for synchronization is that the master stability function (MSF) be strictly negative in each transverse direction \cite{pecora1990synchronization}. With our choice of parameter $\omega$ and  the coupling function $H(\mathbf{x_{i}},\mathbf{x_{j}})=({x_{j}}-{x_{i}},0)^T$, the SL-system belongs to class-I MSF \cite{pecora1990synchronization}, i.e., the synchronization manifold is stable in the interval of
type [$ \beta_{1}, \beta_{2} $]. The $N$ eigen values of $ \overline{G} $ are $ \lambda_1=0 $ and $ \lambda_j=pN $, $j=2,3, . . .,N$. Then the critical range of $k_{A}$ to attain complete synchrony is obtained by the inequality $ \beta_{1} \leq k_A \lambda_j $, $j=2,3,...,N$ which implies, 
\begin{equation} \label{9}
\begin{array}{lcl}
k_{A}\ge\frac{\beta_{1}}{pN}=k_{critical}.
\end{array}
\end{equation}

\begin{figure}[ht]
	\centerline{
		\includegraphics[scale=0.60]{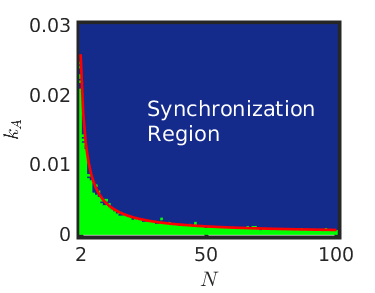}}
	\caption{Synchrony region: The red curve is the derived critical curve plotted using the relation (\ref{9}). The blue dots are the stable synchronization point $(N,k_A)$ at  which $E < 10^{-10}$. The results are accumulated for $50$ independent numerical realizations. The green shaded region is the desynchronization region, with $E \geq 10^{-10}$.}
	\label{picture3}
\end{figure}

\par To verify the relation (\ref{9}), we plot $N-k_{A}$ parameter space in Fig.\ \ref{picture3} where the red solid curve is our analytically found curve  $k_{A}=\frac{\beta_{1}}{pN}$ for $\beta_{1} \simeq 0.0111$. This curve agrees well with our numerically found synchronization region (blue shaded region). In order to numerically study the emergence of synchrony,  we define the synchronization error as
\begin{equation} \label{6}
\begin{array}{lcl}
E = \bigg \langle \frac{\sum_{i,j=1 (i \not=j)}^{N}\sqrt{\left(x_j-x_i\right)^2 + \left(y_j-y_i\right)^2} }{N-1}\bigg \rangle_t,
\end{array}
\end{equation}
where $\langle\cdot\rangle_t$ represents time average obtained over a long time interval (taken as $3 \times 10^5$ time steps here) after initial transient of $3 \times 10^5$ time steps. In our work, complete synchronization corresponds to the state when $E$ becomes less than $10^{-10}$ accounting $50$ independent network realizations. Note that, we find a small fluctuations in the relation (\ref{9}) for finding $k_{critical}$. These fluctuations may occur due to various reasons, including insufficient realizations, approximate choice of the value of $\beta_{1}$, or for the effect of the uniform moving velocity $v$. Even, this small fluctuation may be due to our choice of $p$. Although, for static agent assumption or $v=0$, our chosen $p=1-\frac{\pi \alpha^2}{4g^2}$ works absolutely correctly. But when the agents are moving, the agents do not necessarily maintain this exact relation (\ref{9}). In fact, in deriving the relation (\ref{9}), we use the negativity of the maximum Lyapunov exponent in the transverse direction of the synchronization manifold, which is a necessary condition, not a sufficient one. All these lead to small fluctuations in $k_{critical}$ compared to our derived critical synchronization curve $k_{critical}=\frac{\beta_{1}}{pN}$. Notably, this density dependent threshold for the emergence of synchrony is found to be relevant and is of practical interest, particularly in the studies of the bacterial infection, biofilm formation and bioluminescence where quorum-sensing transition is observed in indirectly coupled systems \cite{camilli2006bacterial,taylor2009dynamical}.
\par This result is quite important due to the fact that if we randomly distribute $N$ agents in the plane $S$ and then set $v=0$ (i.e., when the network becomes static), complete synchronization is not guaranteed depending on their initial placements.  As the connected agents may exhibit complete synchronization depending on suitable coupling strength $k_A$, but the disconnected agents remain isolated and as a result of that they maintain isolated trajectories. But, the mobility of each agent allows them to interact even if they are long-distant, under sufficiently long simulations. These occasional interactions induce complete synchronization among them for appropriate critical coupling strength.

\par Such occasional minimal interaction is found to be beneficial for the emergence of complete synchronization, instead of continuous-time coupling, from the point of view of optimal interactional cost \cite{schroder2015transient,tandon2016synchronizing}. This confined interaction may be useful in case of robotic networks and wireless communication systems, where transmission signals are turned on if the agents are lying sufficiently close to each other and where we will have to inspect if synchronization can still occur.

\begin{figure*}[ht]
	\centerline{
		\includegraphics[scale=0.6]{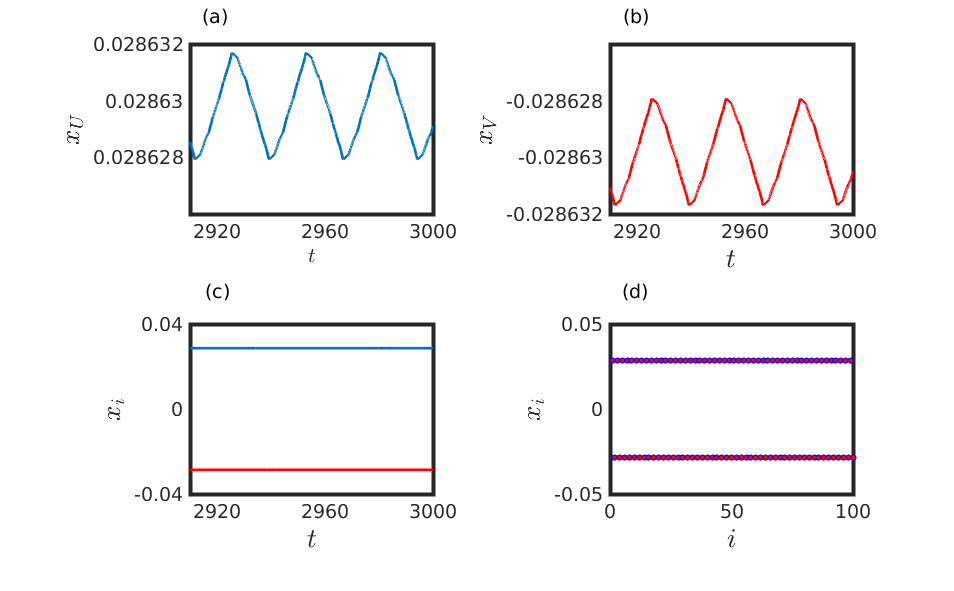}}
	\caption{Cluster Synchronization: The oscillators maintain a small oscillation around (a) $0.02863$ for the set $U$ and (b) $-0.02863$ for the set $V$. All the oscillators of each set $U$ and $V$ converge to a single attractor. (c) The time-series of $x_{i}$ of all $N=100$ oscillators reveal that trajectories converge to two distinct trajectories. (c) These final converged trajectories look like oscillation death states, but the upper panel (a, b) already reveal the existence of small oscillation. (d) Two clusters are clearly observed. Here, the attractive coupling strength $k_A=1.0$.  }
	\label{picture4}
\end{figure*}
\subsection{Effect of global attractive coupling}

\par To analyze the effect of only attractive coupling, one requires to turn-off the repulsive coupling $k_R$. But this is not the only possible way of setting on repulsion free interaction. Instead, one may realize the scenario by setting $\alpha = 0$. This specific choice of $\alpha$ transforms the repulsive matrix $C$ to  null matrix, and attractive matrix in the form

\noindent
$ ~~~~~~~~B=\left(\begin{array}{cccccccc}
0 & 1 & 1 & \cdots & \cdots & \cdots & 1 \\
1 & 0 & 1 & \cdots & \cdots & \cdots & 1 \\
1 & 1 & 0 & \cdots & \cdots & \cdots & 1\\
\cdots & \cdots & \cdots & \cdots & \cdots & \cdots & \cdots\\
\cdots & \cdots & \cdots & \cdots & \cdots & \cdots & \cdots\\
\cdots & \cdots & \cdots & \cdots & \cdots & \cdots & \cdots\\
1 & 1 & 1 & \cdots & \cdots & \cdots & 0
\end{array}\right)$

Note that the matrix $B$ now becomes a static, i.\ e.,\ a time-independent matrix. Under this set-up, one may expect complete synchronization, but for higher values of $k_A$, we can not obtain convergence of those attractors into one attractor. Instead of that, we find incomplete synchronization in the form of cluster synchronization for $k_A \geq 0.19$. We define $U=\{x_{i}|~~ i ~\text{ is odd}\}$ and $V=\{x_{i}|~~i ~ \text{is even}\}$. After the initial transient, the $x_i$ of trajectories of the sets $U$ and $V$ converge respectively to two different trajectories as shown in Figs.\ \ref{picture4}(a, b) for $k_A=1.0$. Both the sets $U$ and $V$ contain $50$ oscillators. The global network evolves into two equipotent subsets of oscillators in which members of the same cluster are synchronized to the same trajectory, but members of different clusters, oscillating with different small amplitudes. Notice that the oscillators that start with negative $x_{i}$ as per relation \eqref{5} end up with positive $x_{i}$ and vice-versa. Figure \ref{picture4}(c) portrays the time series of $x_{i}$ for $i=1,2,\cdots,100$. Although, Fig.\ \ref{picture4}(c) reflects oscillation death like scenario, but Figs.\ \ref{picture4}(a, b) reveal a small oscillation around $0.02863$ for the set $U$ and $-0.02863$ for the set $V$. Figure \ref{picture4}(d) depicts snapshot of the position of the oscillators at a particular time $t$. All $N=100$ oscillators are symmetrically distributed around the origin, the unstable stationary point of the system \eqref{4}.

\par Such partial synchronization may show up in swarms of unmanned autonomous vehicles, power grids and swarms of animals \cite{sorrentino2016complete}. This interesting phenomenon is also hold for random initial conditions choosing from $ [-1, 1] \times [-1, 1]$. But, in that case, the cardinality of $U$ and $V$ vary significantly in each realization. We also find that, if we increase $k_A$ by keeping fixed the initial conditions as per the relation \eqref{5}, still we do not observe complete synchronization for $k_A \geq 0.19$. They still show cluster synchronization for suitable $k_A$. However, for smaller suitable $k_A$, the trajectories exhibit complete synchronization converging to a single periodic attractor. For $N=100$ oscillators, complete synchronization can be obtained for $k_A < 0.19$.

\begin{figure*}[ht]
	\centerline{
		\includegraphics[scale=0.6]{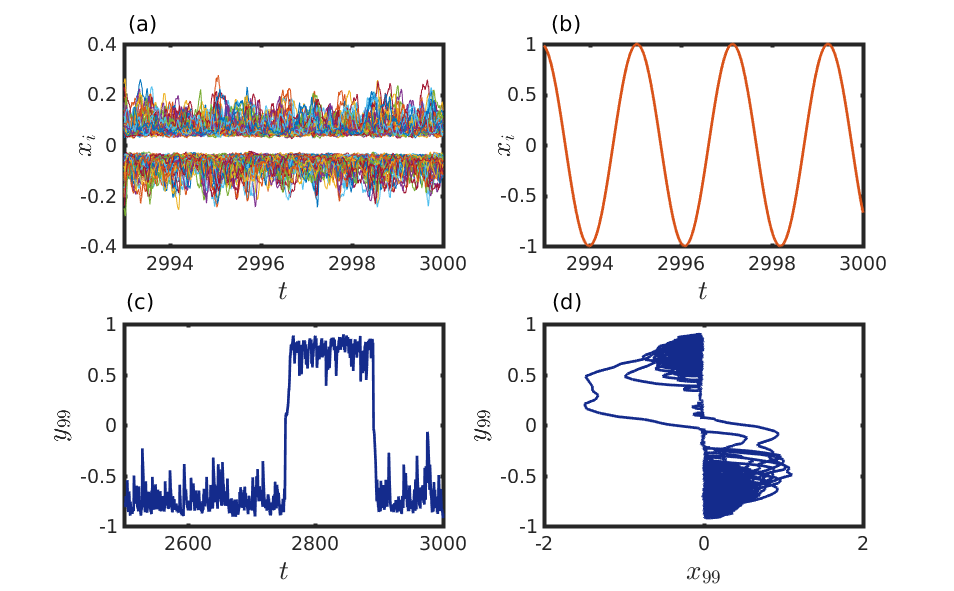}}
	\caption{(a) Inhomogeneous small oscillation: variation of $x_i(t)$ of all the oscillators for $N=100$. (b) Complete synchronization: variation of $x_i(t)$ of $N=51$ oscillators. Other parameters are $k_A=1.0$ and $k_R=-0.10$. The initial conditions are chosen as per the relation (\ref{5}). (c) Attractor switching: Without any loss of generality, the $99$-th oscillator is chosen randomly among $N=100$ oscillators. The time-series depicts the occasional switching of the attractor. (d) The chaotic attractor: A single attractor is shown here to illustrate the attractor switching. Other parameters are $k_A=1.0$ and $k_R=-0.27$. }
	\label{picture5}
\end{figure*}

\subsection{Interplay between attractive and repulsive interactions}

\subsubsection{Inhomogeneous small oscillation}

Now, we consider $\alpha=10$ along with $k_A=1.0$ and $k_R=-0.10$, so that both the couplings can play a decisive role. Under this choice of parameters, we observe the time-series of each oscillator exhibits chaotic behavior. Again, those $x_{i}$ of $N=100$ oscillators distribute equally on both sides of $y_{i}=0$. Similarly, $y_{i}$ of those oscillators are distributed equally on both sides of $x_{i}=0$. One of such scenario is shown in Fig.\ \ref{picture5}(a). It is observed that each of these two groups is not synchronized. They maintain their chaotic small oscillations. The simulations are done using initial conditions of relation (\ref{5}).


\par Recently this type of oscillations is reported by Dixit et al.\cite{dixit2020static}\ for two coupled SL-oscillators under the influence of dynamic interaction.
But, our study is quite different from them. Our network consists of mobile agents, and their relative distance $d_{ij}(t)$ actually decides the mutual interaction type, which is completely different from the study \cite{dixit2020static}. In fact, we found this state for large networks, which help to deny the finite-size effect. Moreover, each split group in our case consists more than $2$-oscillators and those trajectories do not lead to any cluster synchronization.

\par To study the effect of network size $N$,  a detailed numerical study is perceived. We find that there exists a $N_{critical}$, beyond which network shows such inhomogeneous small oscillation. Numerically, we find this $N_{critical}$ is $51$. For $N \geq 52$, such chaotic small oscillation is observed. For $N < 52$, the attractors are converged to a single attractor and the trajectories are exhibiting periodic behavior. In Fig.\ \ref{picture5}(b), we plot the $x_{i}$'s of all $N=51$ oscillators, which are oscillating synchronously.

\begin{figure}[ht]
	\centerline{
		\includegraphics[scale=0.4]{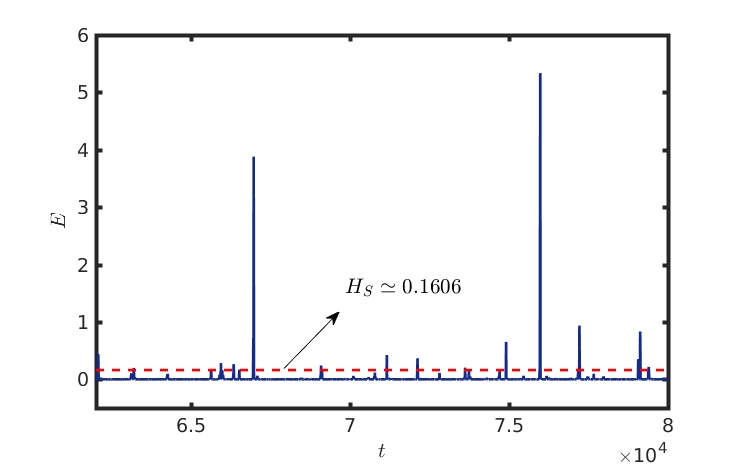}}
	\caption{On-off intermittency: The red dashed horizontal line is the extreme event indicating threshold $H_S=m+ 8\sigma$, where $m$ is the sample mean and $\sigma$ is the standard deviation of the whole data. The error dynamics, $E$ shows irregular and rare switching from the zero value to non-zero values. The parameters are $N=100$, $k_A=1.0$, $\alpha=g=10.0$, $v=2.0$ and $k_R=-0.44$.   }
	\label{picture6}
\end{figure}

To scrutinize the effect of $k_R$, we fix the attractive coupling $k_A=1.0$, and slowly increase the absolute magnitude of $k_R$. We observe that with increment of $k_R$, the amplitude of the oscillators increase. After a certain value of $k_R$, the attractors of one group, which are initially separated by the axes (See Fig.\ \ref{picture5}(a)), are crossing the axes frequently. The trajectory of $i$-th oscillator crosses the axes and joins the other group for sometimes, before it comes back to its original group. To illustrate this, we choose randomly an agent from $N=100$ oscillators, and plot its time series in Fig.\ \ref{picture5}(c). The corresponding attractor is also shown in Fig.\ \ref{picture5}(d) to demonstrate the chaotic switching behavior of the attractor.  Interestingly, further slight increment of $k_R$ leads to complete synchronization. At $k_R=-0.28$, all $N=100$ oscillators oscillate synchronously and re-gain their isolated periodic dynamics. They oscillate in the limit cycle regime from $[-1,1]$. 

\subsubsection{Extreme events}
\par Further decrement of $k_R$ (for fixed $k_A=1.0$) destabilizes the synchronization manifold and after a certain value of $k_R$, the error dynamics becomes intermittent. Specifically, for most of the time the synchronization error $E$ (cf. Eq.\ (\ref{6})) remains near zero (lies within the interval $[10^{-10}, 10^{-6}]$), but occasionally this $E$ becomes non-zero. For a narrow range of $k_R$ with fixed $k_A=1.0$, we find this intermittent error trajectories exhibiting extremely large values. 
To reveal this issue, we plot variation of synchronization error $E$ in Fig.\ \ref{picture6}, which is accumulated over a sufficiently long time interval.
For this entire data, we calculate the sample mean $m = 0.0011$ and the standard deviation $\sigma = 0.0199$. We then define a extreme event threshold $H_S=m+ 8\sigma \simeq 0.1606$, which is plotted over the error dynamics with horizontal dashed line in Fig.\ \ref{picture6}. In this particular accumulated data of $E$, we find $2570$ data points having value of $E$, which is greater than $H_S$. Thus, the probability of $\{E > H_{S}\}$ is $0.0008567$. However, this probability varies with each realization, and for our choice of $H_S=m+ 8\sigma$, Cantelli's inequality \cite{cantelli1929sui,ghosh2002probability} yields the upper bound for each realization as

\begin{equation} \label{prob}
\begin{array}{lcl}
P(E \geq H_{S}) \leq \frac{1}{65}.
\end{array}
\end{equation}

\par In the existing literatures \cite{heagy1995desynchronization,platt1993off}, this type of intermittency is known as \textit{on-off intermittency}. This type of intermittent error, the trajectories experiencing non-uniform, uncertain extensive expeditions from the synchronous manifold and can be considered as an appropriate candidate for extreme events. Earlier, local instability of synchronization manifold due to on-off intermittency causing extreme events in a pair of coupled chaotic electronic circuits is reported \cite{cavalcante2013predictability}. Also, Nag Chowdhury et al.\ \cite{chowdhury2019synchronization} found a similar mechanism for the generation of extreme events in a mobile network of chaotic oscillators. 

\begin{figure}[ht]
	\centerline{
		\includegraphics[scale=0.3]{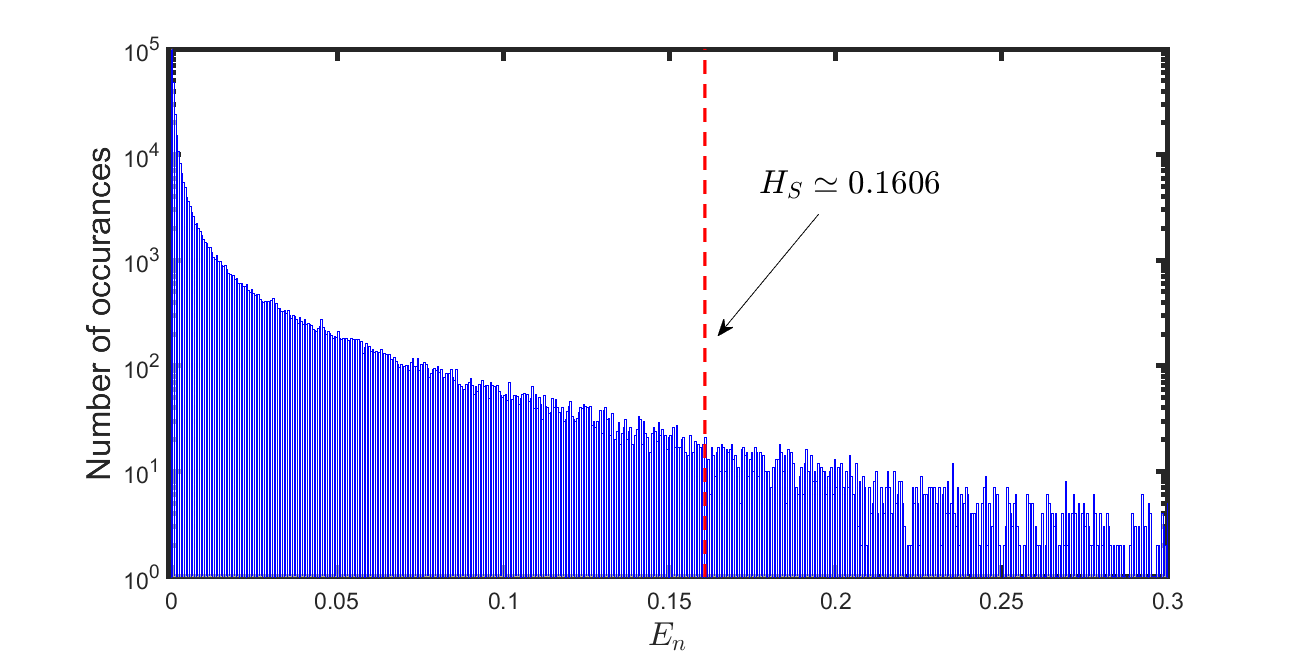}}
	\caption{Non-Gaussian distribution of $E_n$:  Local maxima $E_{n}$ of $E$ is accumulated and histogram of them is plotted in semilog scale. This histogram resemblances with L-shaped PDF exhibiting non-Gaussian distribution. The red dashed vertical line is the significant height $H_{S}=m+8\sigma$. The parameters are $N=100$, $k_A=1.0$, $\alpha=g=10$, $v=2.0$ and $k_R=-0.44$.   }
	\label{picture7}
\end{figure}

\begin{figure*}[ht]
	\centerline{
		\includegraphics[scale=0.40]{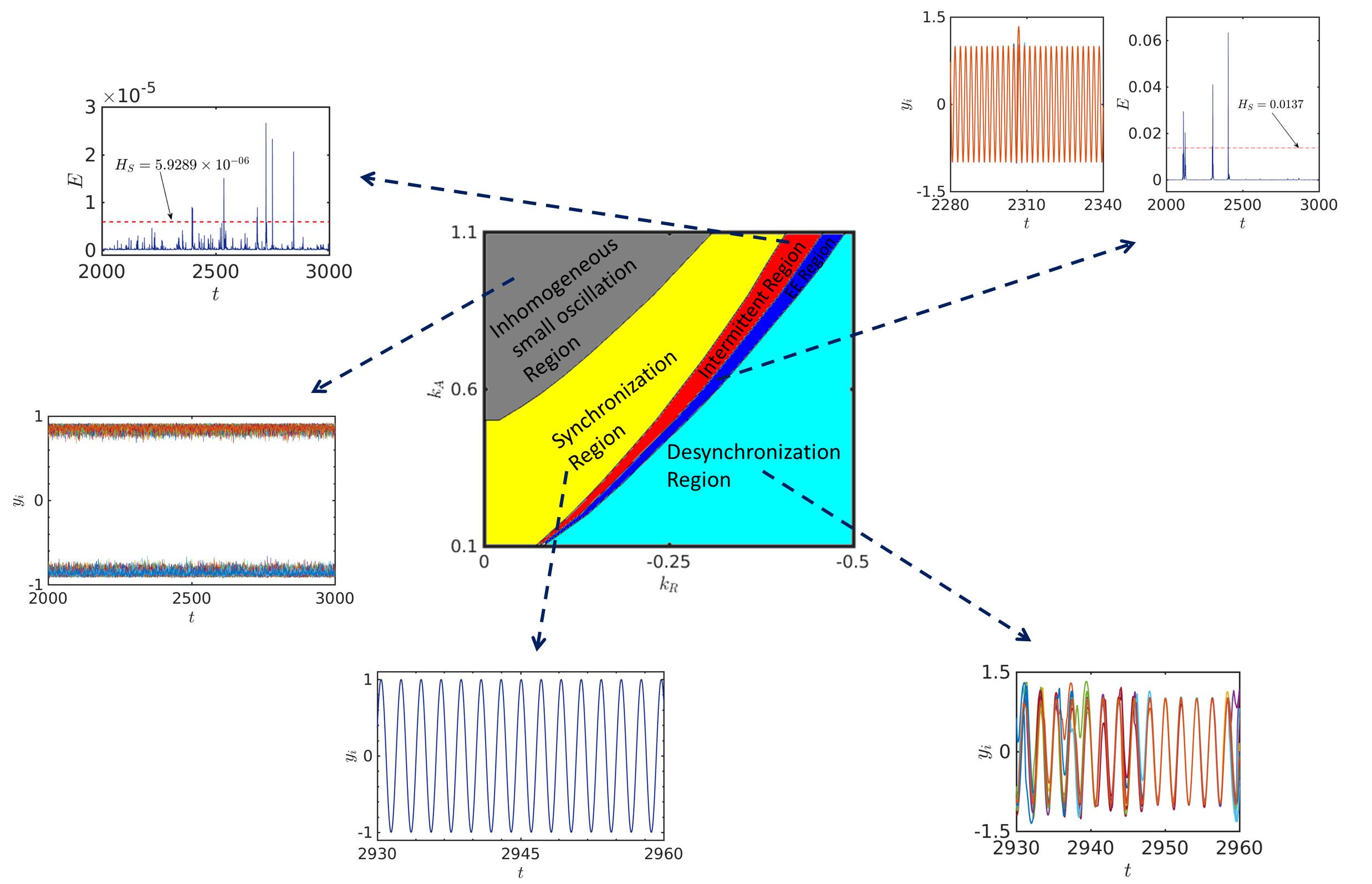}}
	\caption{Parameter Space $k_R-k_A$: The yellow and gray regions respectively correspond to the states $E < 10^{-10} \simeq 0$ (i.e., complete synchrony) and the inhomogeneous small oscillation. The blue area stands for the extreme events whereas the red zone represents the state where the error $E$ becomes intermittent but $H_{S} \leq 0.01$. Finally, the area in cyan corresponds to the desynchronization state. Other parameters are $N=100$, $g=\alpha=10$ and $v=2.0$. Here, $100$ independent realizations are  used to obtain this parameter space.}
	\label{picture9}
\end{figure*}

\par To further resolve whether the spikes in Fig.\ \ref{picture6} qualify for extreme events or not, we determine the local maximum values $E_{n}$ of $E$ and plot the histogram for the event sizes $E_{n}$ in Fig.\ \ref{picture7}. Note that the length of the collected time series is sufficient enough so that due to statistical regularity, inclusion of new sample does not affect the structure of the histogram. Pisarchik et al. \cite{pisarchik2011rogue} pointed out a special characteristic of extreme events, which is the L-shaped probability density function (PDF) of $E_{n}$ in the semi-log scale. In spirit of all these facts, we define extreme events in this article as follows,

\begin{enumerate}
	\item These short lasting events are recurrent, aperiodic and of different amplitudes higher than the significant height $H_S=m+8\sigma$. In order to further eliminate the small amplitude events, another restriction is maintained, $H_S > 0.01$ is taken along with $H_S=m+8\sigma$ \cite{chowdhury2019synchronization,pisarchik2011rogue}.
	
	\item They appear much more often than Gaussian statistics and they are unpredictable almost surely with respect to time, where an event is defined as almost surely if the set of possible exceptions may be non-empty but has zero Lebesgue measure \cite{akhmediev2010editorial}.
	
	\item The appearance of these large events would have very small probability (though the exact measure of small needs to be quantified) \cite{lucarini2016extremes}. 
\end{enumerate}

\par Clearly, the histogram in Fig.\ (\ref{picture7}) is an approximate representation of the distribution $f(x)$ of the numerical data. In the next, we define a return interval $k$. Mathematically, a return interval $k$ occurs if $X_{i} > H_{S}$ and $X_{i+k} > H_{S}$, but $X_{j} \leq H_{S}$ for $i < j < i+k$, where $X_{i}$ is the value of the observable $X$ at the $i$-th step. Suppose, $W_{H_{S}}$ is the number of total return intervals. So, we have $k_{i}$ for $i=1,2,\cdots,W_{H_{S}}$ such that not necessarily all the $k_{i}$ are distinct. Clearly, if $k_{i}$ is short then a cluster of accumulated extreme events can be observed. In contrast, a large value of return interval definitely portrays few occurrences of extreme events. Also, let $k_{1}+k_{2}+\cdots+k_{W_{H_{S}}}=W$, then the mean return interval is

\begin{equation} \label{10}
\begin{array}{lcl}
R_{H_{S}}=\frac{1}{W_{H_S}}\sum_{i=1}^{W_{H_{S}}} k_{i}=\frac{W}{W_{H_S}}.
\end{array}
\end{equation}

This relation (\ref{10}) clearly indicates that the mean return interval $R_{H_{S}}$ is relatively short if the occurrence of extreme events is frequent (i.e.,\ $W_{H_{S}}$ is high). Similarly, if $W_{H_{S}}$ is small then the mean return interval is relatively high. But, this number of return intervals $W_{H_{S}}$ is dependent on various factors and thus it varies on realizations. Hence, relation (\ref{10}) gives different average return interval $R_{H_{S}}$ based on different realizations. Using time series analogous of Kac’s Lemma \cite{kac1947notion,altmann2005recurrence,eichner2007statistics}, $R_{H_{S}}$ can be described in terms of the tail of the normalized distribution density $f(x)$ as
\begin{equation} \label{11}
\begin{array}{lcl}
R_{H_{S}}=\frac{W}{W_{H_S}} \simeq \frac{1}{\int_{H_{S}}^{\infty} f(y)dy}.
\end{array}
\end{equation}
Thus, there exists a one-one relation between the chosen threshold $H_S$ and the mean return interval $R_{H_{S}}$, which is solely determined by the normalized distribution $f(x)$.

\subsubsection{Different dynamical states in the $k_A - k_R$ coupling parameter space}

\par For a complete understanding of interplay between attractive and repulsive interactions (i.e., for $\alpha\ne 0$), a two dimensional $k_{A}-k_{R}$ parameter space is presented in Fig.\ \ref{picture9}. In this figure, we vary $k_A$ from $[0.1,1.1]$ whereas $k_R$ is varied within $[0.0,-0.50]$.  Initially, in absence of repulsive coupling, i.e.,\ for $k_R = 0.0$, complete synchronization is observed for suitable choices of $k_A$ where $E < 10^{-10}$. As per this Fig.\ \ref{picture9}, whenever $k_A$ reaches to $0.5$, then the introduction of small $k_R < 0.0$ may destroy the limit cycle behavior of the oscillators and their synchronized behavior is also lost. The reflected region of inhomogeneous small oscillation enlarges with further increment of positive coupling $k_A$. Interestingly, for a fixed $k_A \geq 0.5$, the decrement of $k_R$ helps to restore the periodic behavior of the attractors. With suitable choices of $k_R$ for fixed $k_A \geq 0.5$, this transition from inhomogeneous small oscillation to complete synchronized state occurs through the path of attractor switching, as already shown in Fig.\ \ref{picture5}. The attractors are quenched and with further decrement of $k_R$ for appropriate $k_A$, the trajectories switch between the co-existing attractors and finally regain their respective periodic synchronized attractor.
\par When $k_R$ is further decreased, the synchronization error $E$ becomes intermittent and reveals occasional away journey from the synchronization manifold. This regime is depicted through the red and blue regions in the Fig.\ (\ref{picture9}). In order to distinguish between these red and blue regions, we calculate the significant height $H_{S}$. When $H_{S}$ lies within $(10^{-6},0.01]$, we label those states as the intermittent region (in red). For $H_{S} > 0.01$, we define those states as extreme events which obey additional few clauses mentioned earlier. While in complete synchronization, all the attractors converge into a single attractor. But, during the appearance of extreme events, most of the time all the trajectories evolve over a converged single trajectory, but occasionally due to the local repulsion (of suitable strength $k_R$) few trajectories leave the converged attractor and follow their own path for a relatively short time. Hence, during those time, the error trajectory leaves the equilibrium point $E=0.0$ (the synchronization manifold) and exhibit a large excursion exceeding $H_{S}=m+8\sigma$. This choice of $H_S$ is influenced by the Ref.\ \cite{chowdhury2019synchronization}, where the authors gave analytical logic behind the choice of such extreme event indicating threshold. The second variables $y_{i}$ of all $N=100$ oscillators are plotted over a short time-interval and the corresponding error dynamics are also presented through an arrow in Fig.\ \ref{picture9}. As expected, if $k_R$ is further decreased while keeping $k_A$ fixed, desynchrony states are observed.

\begin{figure}[ht]
	\centerline{
		\includegraphics[scale=0.35]{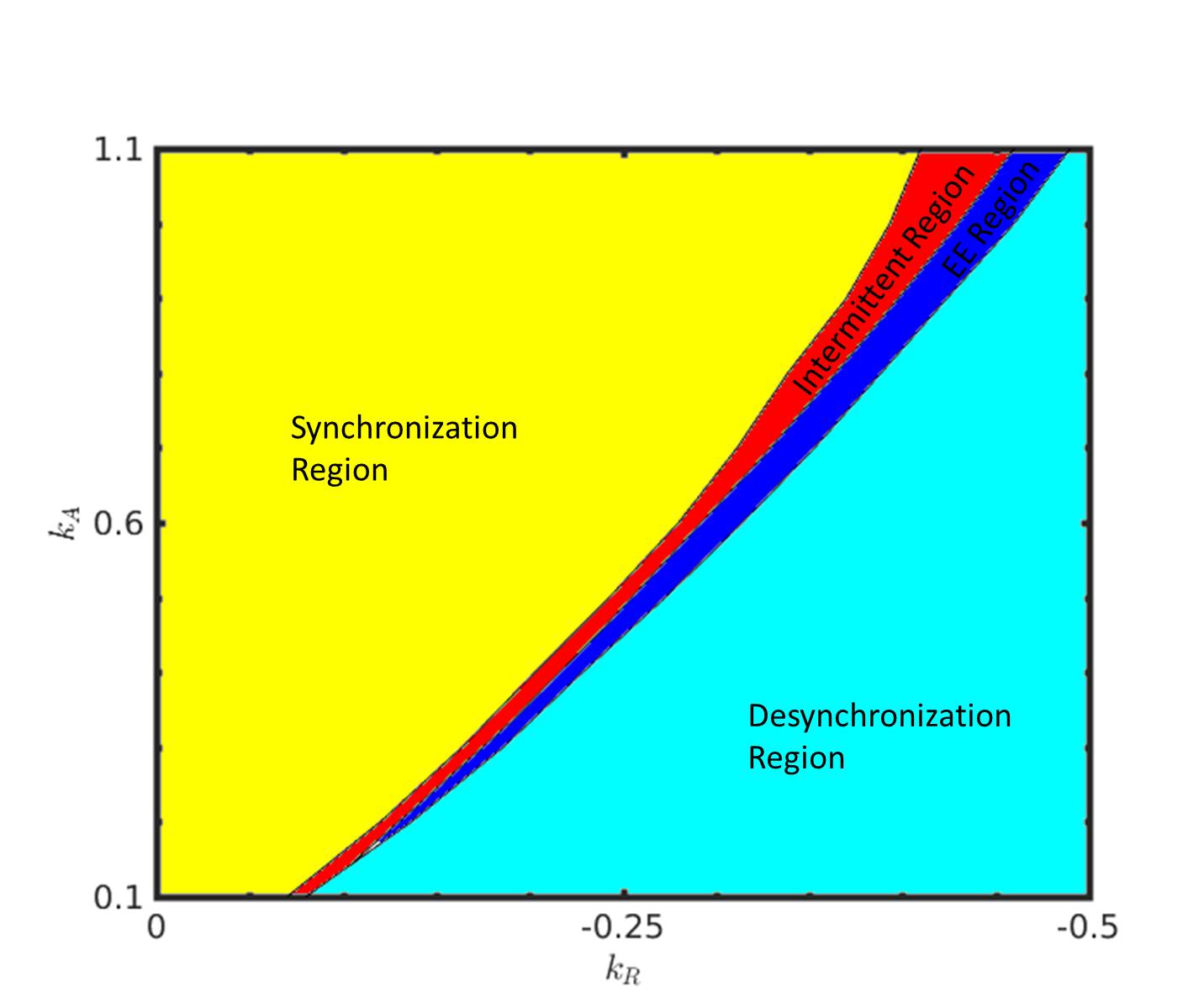}}
	\caption{Phase diagram in the $k_R-k_A$ parameter plane: The yellow region corresponds to the states of complete synchronization. The blue area stands for the extreme events whereas the red zone represents the state where the error $E$ becomes intermittent, but $H_{S} \leq 0.01$. Finally, the area in cyan corresponds to the desynchronization state. Other parameters are $N=100$, $g=\alpha=10$ and $v=2.0$. Here, symmetry-preserving vector coupling among the local dynamical units is considered in which $100$ independent realizations are used to obtain this parameter space.}
	\label{picture10}
\end{figure}

\section{Symmetry-preserving coupling} \label{Symmetry-preserving coupling}

In this section, we deal with the scenario in which the local dynamical units, i.e., the SL-oscillators are coupled through both the variables so that the coupling becomes (rotational) symmetry-preserving. Our main emphasis will be to examine how this change affects the collective dynamics of the networked system. We depict the dynamical outcomes through Fig.\ \ref{picture10} in which the phase diagram in the $k_R-k_A$ parameter plane is portrayed. As can be seen from the figure, the presence of only attractive coupling readily induces complete synchronization in the system. Moreover, in contrast to our earlier observation for symmetry-breaking coupling (cf. Fig.\ \ref{picture9}), this synchrony sustains as $k_A$ increases. This is mainly because now the coupling function does not allow to break the symmetry yielding inhomogeneous small oscillation. On the other hand, switching the repulsive coupling strength $k_R$ on, one observes occasional jumps in the error dynamics (for suitable strengths of couplings) that indicates the appearance of intermittent states. More importantly, further increment of $k_R$ leads to the intermittent states that satisfy the criteria of being considered as extreme events. This result demonstrates that our observations of complete synchronization or the extreme events on the considered network set-up is not limited to a specific choice of symmetry-breaking coupling function.

\section{Discussion and Conclusion} \label{Discussion and Conclusion}

\par In this study, we have considered Stuart-Landau oscillators on top of mobile agents moving in a finite region of two-dimensional plane. Due to the spatial movement of the agents, the relative distance between any two agents always varies with time. Using this spatial distance between the agents, two competing interactions are introduced. Whenever the agents stay inside a closed ball of radius $\alpha$, the oscillators are repulsively coupled bidirectionally through diffusive linear coupling. On the contrary, whenever the mobile agents lie beyond the closed communication ball of radius $\alpha$, they are coupled through attractive coupling. Under this set-up of the networked system, we have encountered diverse collective behaviors and provided rigorous investigation of each of them. 

\par Firstly, we have addressed the effect of sole attractive interaction among the dynamical agents. In order to accomplish this, we first set the repulsive strength to zero for which complete synchronization is found for $\alpha=10$. This result is quite interesting as for a static time-independent network, complete synchronization is not possible. Mobility of the agents play the decisive role to converge all the attractors to a single one. Analytically, the critical attractive coupling strength for synchrony is calculated, which exhibit excellent agreement with the numerical results. To investigate further the sole influence of attractive coupling, the repulsive matrix is set to be a null matrix by keeping $\alpha = 0$ fixed. This different strategy leads to complete synchronization for suitable choice of $k_A$ and cluster synchronization beyond a certain value of $k_A$. 

\par Next we dealt with our prime issue of coexisting attraction and repulsion. The coexistence of such competing interactions can demolish the synchronized behavior of the whole system. Under suitable choices of both coupling strength, inhomogeneous small oscillation is perceived. In this state, each oscillator maintains chaotic behavior though we set initially each oscillator into limit cycle regime. Further reduction of repulsive strength helps to regain their periodic behavior through the path of attractor switching. Moreover, further lessening leaves the error dynamics intermittent. This irregular away journey of error trajectory from the synchronization manifold gives rise to infrequent large deviated events. Using few characterizations, we confirmed these states as extreme events. The mean return interval of these extreme events is approximated using time series analogs of Kac's lemma in relation (\ref{11}). This relation indicates the dependency of average return interval $R_{H_{S}}$ on the threshold $H_S$. The larger $R_{H_{S}}$ clearly corresponds to larger $H_S$ and vice-versa. The route for generating such large amplitude events is also addressed in terms of on-off intermittency. An upper bound for the probability of occurrence of extreme events is calculated in relation (\ref{prob}) depending on the choice of $H_{S}$.

\par Further, to elucidate the mechanism behind inhomogeneous small oscillation of those oscillators, vector coupling is introduced. This coupling helps to preserve the rotational symmetry of SL-oscillators. Although, this vector coupling yields behavior like complete synchrony and extreme events, but it eliminates the appearance of such small oscillation, which is occurred during symmetry-breaking coupling.

\par We have been able to portray the whole scenario of all these emerging collective behaviors in a two-dimensional parameter space of the competing coupling strengths. This parameter space manifests the fact that by tuning any one of the coupling strength, one can avoid such devastating extreme events. This fact can be used as one of the influential strategy for controlling extreme events. Such controlled usage of antibiotics \cite{chen2018social,chen2019imperfect} already found to be helpful in avoiding worsening medical cost and mortality especially for life-threatening bacteria infections.

\par  In this article, we analyzed the interplay of positive and negative interactions in coupled identical limit cycle oscillators, by considering SL system on top of each mobile agent. An important direction of future generalization may include networks of chaotic oscillators, which is where our approach might make a difference. The chaotic dynamics \cite{rossler1976equation,yan2019design,he2020fractional,yan2020multistability} are expected to reveal an even wider spectrum of interesting dynamical states. This remains the core interesting avenue for future work.

\section*{Acknowledgements}

SNC and DG were supported by Department of Science and Technology, Government of India (Project No. EMR/2016/001039).  SNC would also like to thank Physics and Applied Mathematics Unit of Indian Statistical Institute, Kolkata for their support during the pandemic COVID-19.


\begin{thebibliography}{10}
	\providecommand{\url}[1]{#1}
	\csname url@samestyle\endcsname
	\providecommand{\newblock}{\relax}
	\providecommand{\bibinfo}[2]{#2}
	\providecommand{\BIBentrySTDinterwordspacing}{\spaceskip=0pt\relax}
	\providecommand{\BIBentryALTinterwordstretchfactor}{4}
	\providecommand{\BIBentryALTinterwordspacing}{\spaceskip=\fontdimen2\font plus
		\BIBentryALTinterwordstretchfactor\fontdimen3\font minus
		\fontdimen4\font\relax}
	\providecommand{\BIBforeignlanguage}[2]{{%
			\expandafter\ifx\csname l@#1\endcsname\relax
			\typeout{** WARNING: IEEEtran.bst: No hyphenation pattern has been}%
			\typeout{** loaded for the language `#1'. Using the pattern for}%
			\typeout{** the default language instead.}%
			\else
			\language=\csname l@#1\endcsname
			\fi
			#2}}
	\providecommand{\BIBdecl}{\relax}
	\BIBdecl
	
	\bibitem{newman_10}
	M.~E.~J. Newman, \emph{Networks: An Introduction}.\hskip 1em plus 0.5em minus
	0.4em\relax Oxford, U.K.: Oxford University Press, 2010.
	
	\bibitem{barabasi2016network}
	A.-L. Barab{\'a}si \emph{et~al.}, \emph{Network science}.\hskip 1em plus 0.5em
	minus 0.4em\relax Cambridge university press, 2016.
	
	\bibitem{albert2002statistical}
	R.~Albert and A.-L. Barab{\'a}si, ``Statistical mechanics of complex
	networks,'' \emph{Reviews of Modern Physics}, vol.~74, no.~1, p.~47, 2002.
	
	\bibitem{boccaletti2006complex}
	S.~Boccaletti, V.~Latora, Y.~Moreno, M.~Chavez, and D.-U. Hwang, ``Complex
	networks: Structure and dynamics,'' \emph{Phys. Rep.}, vol. 424, no. 4-5, pp.
	175--308, 2006.
	
	\bibitem{arenas2008synchronization}
	A.~Arenas, A.~D{\'\i}az-Guilera, J.~Kurths, Y.~Moreno, and C.~Zhou,
	``Synchronization in complex networks,'' \emph{Phys. Rep.}, vol. 469, no.~3,
	pp. 93--153, 2008.
	
	\bibitem{rakshit2020invariance}
	S.~Rakshit, B.~K. Bera, and D.~Ghosh, ``Invariance and stability conditions of
	interlayer synchronization manifold,'' \emph{Physical Review E}, vol. 101,
	no.~1, p. 012308, 2020.
	
	\bibitem{wang2002synchronization}
	X.~F. Wang and G.~Chen, ``Synchronization in scale-free dynamical networks:
	robustness and fragility,'' \emph{IEEE Transactions on Circuits and Systems
		I: Fundamental Theory and Applications}, vol.~49, no.~1, pp. 54--62, 2002.
	
	\bibitem{vaidyanathan2014adaptive}
	S.~Vaidyanathan, C.~Volos, V.-T. Pham, K.~Madhavan, and B.~A. Idowu, ``Adaptive
	backstepping control, synchronization and circuit simulation of a 3-d novel
	jerk chaotic system with two hyperbolic sinusoidal nonlinearities,''
	\emph{Archives of Control Sciences}, vol.~24, no.~3, 2014.
	
	\bibitem{rakshit2020intralayer}
	S.~Rakshit, B.~K. Bera, E.~M. Bollt, and D.~Ghosh, ``Intralayer synchronization
	in evolving multiplex hypernetworks: Analytical approach,'' \emph{SIAM
		Journal on Applied Dynamical Systems}, vol.~19, no.~2, pp. 918--963, 2020.
	
	\bibitem{vaidyanathan2015analysis}
	S.~Vaidyanathan, C.~Volos, V.-T. Pham, and K.~Madhavan, ``Analysis, adaptive
	control and synchronization of a novel 4-d hyperchaotic hyperjerk system and
	its spice implementation,'' \emph{Archives of Control Sciences}, vol.~25,
	no.~1, 2015.
	
	\bibitem{wu1996conjecture}
	C.~W. Wu and L.~O. Chua, ``On a conjecture regarding the synchronization in an
	array of linearly coupled dynamical systems,'' \emph{IEEE Transactions on
		Circuits and Systems I: Fundamental Theory and Applications}, vol.~43, no.~2,
	pp. 161--165, 1996.
	
	\bibitem{rakshit2018synchronization}
	S.~Rakshit, B.~K. Bera, and D.~Ghosh, ``Synchronization in a temporal multiplex
	neuronal hypernetwork,'' \emph{Physical Review E}, vol.~98, no.~3, p. 032305,
	2018.
	
	\bibitem{kerr2002local}
	B.~Kerr, M.~A. Riley, M.~W. Feldman, and B.~J. Bohannan, ``Local dispersal
	promotes biodiversity in a real-life game of rock--paper--scissors,''
	\emph{Nature}, vol. 418, no. 6894, pp. 171--174, 2002.
	
	\bibitem{reichenbach2007mobility}
	T.~Reichenbach, M.~Mobilia, and E.~Frey, ``Mobility promotes and jeopardizes
	biodiversity in rock--paper--scissors games,'' \emph{Nature}, vol. 448, no.
	7157, pp. 1046--1049, 2007.
	
	\bibitem{song2010limits}
	C.~Song, Z.~Qu, N.~Blumm, and A.-L. Barab{\'a}si, ``Limits of predictability in
	human mobility,'' \emph{Science}, vol. 327, no. 5968, pp. 1018--1021, 2010.
	
	\bibitem{buscarino2016interaction}
	A.~Buscarino, L.~Fortuna, M.~Frasca, and S.~Frisenna, ``Interaction between
	synchronization and motion in a system of mobile agents,'' \emph{Chaos: An
		Interdisciplinary Journal of Nonlinear Science}, vol.~26, no.~11, p. 116302,
	2016.
	
	\bibitem{stilwell2006sufficient}
	D.~J. Stilwell, E.~M. Bollt, and D.~G. Roberson, ``Sufficient conditions for
	fast switching synchronization in time-varying network topologies,''
	\emph{SIAM Journal on Applied Dynamical Systems}, vol.~5, no.~1, pp.
	140--156, 2006.
	
	\bibitem{nag2020cooperation}
	S.~Nag~Chowdhury, S.~Kundu, M.~Duh, M.~Perc, and D.~Ghosh, ``Cooperation on
	interdependent networks by means of migration and stochastic imitation,''
	\emph{Entropy}, vol.~22, no.~4, p. 485, 2020.
	
	\bibitem{frasca2008synchronization}
	M.~Frasca, A.~Buscarino, A.~Rizzo, L.~Fortuna, and S.~Boccaletti,
	``Synchronization of moving chaotic agents,'' \emph{Physical Review Letters},
	vol. 100, no.~4, p. 044102, 2008.
	
	\bibitem{fujiwara2011synchronization}
	N.~Fujiwara, J.~Kurths, and A.~D{\'\i}az-Guilera, ``Synchronization in networks
	of mobile oscillators,'' \emph{Physical Review E}, vol.~83, no.~2, p. 025101,
	2011.
	
	\bibitem{uriu2013dynamics}
	K.~Uriu, S.~Ares, A.~C. Oates, and L.~G. Morelli, ``Dynamics of mobile coupled
	phase oscillators,'' \emph{Physical Review E}, vol.~87, no.~3, p. 032911,
	2013.
	
	\bibitem{majhi2017synchronization}
	S.~Majhi and D.~Ghosh, ``Synchronization of moving oscillators in three
	dimensional space,'' \emph{Chaos: An Interdisciplinary Journal of Nonlinear
		Science}, vol.~27, no.~5, p. 053115, 2017.
	
	\bibitem{uriu2010random}
	K.~Uriu, Y.~Morishita, and Y.~Iwasa, ``Random cell movement promotes
	synchronization of the segmentation clock,'' \emph{Proceedings of the
		National Academy of Sciences}, vol. 107, no.~11, pp. 4979--4984, 2010.
	
	\bibitem{chowdhury2019convergence}
	S.~N. Chowdhury, S.~Majhi, D.~Ghosh, and A.~Prasad, ``Convergence of chaotic
	attractors due to interaction based on closeness,'' \emph{Physics Letters A},
	vol. 383, no.~35, p. 125997, 2019.
	
	\bibitem{porfiri2006random}
	M.~Porfiri, D.~J. Stilwell, E.~M. Bollt, and J.~D. Skufca, ``Random talk:
	Random walk and synchronizability in a moving neighborhood network,''
	\emph{Physica D: Nonlinear Phenomena}, vol. 224, no. 1-2, pp. 102--113, 2006.
	
	\bibitem{chowdhury2019synchronization1}
	S.~N. Chowdhury and D.~Ghosh, ``Synchronization in dynamic network using
	threshold control approach,'' \emph{EPL (Europhysics Letters)}, vol. 125,
	no.~1, p. 10011, 2019.
	
	\bibitem{kim2013emergence}
	B.~Kim, Y.~Do, and Y.-C. Lai, ``Emergence and scaling of synchronization in
	moving-agent networks with restrictive interactions,'' \emph{Physical Review
		E}, vol.~88, no.~4, p. 042818, 2013.
	
	\bibitem{majhi2019emergence}
	S.~Majhi, D.~Ghosh, and J.~Kurths, ``Emergence of synchronization in multiplex
	networks of mobile r{\"o}ssler oscillators,'' \emph{Physical Review E},
	vol.~99, no.~1, p. 012308, 2019.
	
	\bibitem{bera2016emergence}
	B.~K. Bera, C.~Hens, and D.~Ghosh, ``Emergence of amplitude death scenario in a
	network of oscillators under repulsive delay interaction,'' \emph{Physics
		Letters A}, vol. 380, no. 31-32, pp. 2366--2373, 2016.
	
	\bibitem{mishra2015chimeralike}
	A.~Mishra, C.~Hens, M.~Bose, P.~K. Roy, and S.~K. Dana, ``Chimeralike states in
	a network of oscillators under attractive and repulsive global coupling,''
	\emph{Physical Review E}, vol.~92, no.~6, p. 062920, 2015.
	
	\bibitem{wang2011synchronous}
	Q.~Wang, G.~Chen, and M.~Perc, ``Synchronous bursts on scale-free neuronal
	networks with attractive and repulsive coupling,'' \emph{PLoS one}, vol.~6,
	no.~1, 2011.
	
	\bibitem{hens2014diverse}
	C.~Hens, P.~Pal, S.~K. Bhowmick, P.~K. Roy, A.~Sen, and S.~K. Dana, ``Diverse
	routes of transition from amplitude to oscillation death in coupled
	oscillators under additional repulsive links,'' \emph{Physical Review E},
	vol.~89, no.~3, p. 032901, 2014.
	
	\bibitem{tian2009interlinking}
	X.-J. Tian, X.-P. Zhang, F.~Liu, and W.~Wang, ``Interlinking positive and
	negative feedback loops creates a tunable motif in gene regulatory
	networks,'' \emph{Physical Review E}, vol.~80, no.~1, p. 011926, 2009.
	
	\bibitem{majhi2019solitary}
	S.~Majhi, T.~Kapitaniak, and D.~Ghosh, ``Solitary states in multiplex networks
	owing to competing interactions,'' \emph{Chaos: An Interdisciplinary Journal
		of Nonlinear Science}, vol.~29, no.~1, p. 013108, 2019.
	
	\bibitem{maistrenko2014solitary}
	Y.~Maistrenko, B.~Penkovsky, and M.~Rosenblum, ``Solitary state at the edge of
	synchrony in ensembles with attractive and repulsive interactions,''
	\emph{Physical Review E}, vol.~89, no.~6, p. 060901, 2014.
	
	\bibitem{sathiyadevi2018distinct}
	K.~Sathiyadevi, V.~Chandrasekar, D.~Senthilkumar, and M.~Lakshmanan, ``Distinct
	collective states due to trade-off between attractive and repulsive
	couplings,'' \emph{Physical Review E}, vol.~97, no.~3, p. 032207, 2018.
	
	\bibitem{sathiyadevi2018stable}
	K.~Sathiyadevi, V.~Chandrasekar, and D.~Senthilkumar, ``Stable amplitude
	chimera in a network of coupled stuart-landau oscillators,'' \emph{Physical
		Review E}, vol.~98, no.~3, p. 032301, 2018.
	
	\bibitem{hong2011kuramoto}
	H.~Hong and S.~H. Strogatz, ``Kuramoto model of coupled oscillators with
	positive and negative coupling parameters: an example of conformist and
	contrarian oscillators,'' \emph{Physical Review Letters}, vol. 106, no.~5, p.
	054102, 2011.
	
	\bibitem{vaz2011synchronisation}
	T.~Vaz~Martins and R.~Toral, ``Synchronisation induced by repulsive
	interactions in a system of van der pol oscillators,'' \emph{Progress of
		Theoretical Physics}, vol. 126, no.~3, pp. 353--368, 2011.
	
	\bibitem{leyva2006sparse}
	I.~Leyva, I.~Sendina-Nadal, J.~Almendral, and M.~Sanju{\'a}n, ``Sparse
	repulsive coupling enhances synchronization in complex networks,''
	\emph{Physical Review E}, vol.~74, no.~5, p. 056112, 2006.
	
	\bibitem{mishra2018dragon}
	A.~Mishra, S.~Saha, M.~Vigneshwaran, P.~Pal, T.~Kapitaniak, and S.~K. Dana,
	``Dragon-king-like extreme events in coupled bursting neurons,''
	\emph{Physical Review E}, vol.~97, no.~6, p. 062311, 2018.
	
	\bibitem{inoue2010dynamics}
	M.~Inoue and K.~Kaneko, ``Dynamics of coupled adaptive elements: Bursting and
	intermittent oscillations generated by frustration in networks,''
	\emph{Physical Review E}, vol.~81, no.~2, p. 026203, 2010.
	
	\bibitem{ottino2018volcano}
	B.~Ottino-L{\"o}ffler and S.~H. Strogatz, ``Volcano transition in a solvable
	model of frustrated oscillators,'' \emph{Physical Review Letters}, vol. 120,
	no.~26, p. 264102, 2018.
	
	\bibitem{bacelar2014exploring}
	F.~S. Bacelar, J.~M. Calabrese, and E.~Hern{\'a}ndez-Garc{\'\i}a, ``Exploring
	the tug of war between positive and negative interactions among savanna
	trees: Competition, dispersal, and protection from fire,'' \emph{Ecological
		Complexity}, vol.~17, pp. 140--148, 2014.
	
	\bibitem{szell2010multirelational}
	M.~Szell, R.~Lambiotte, and S.~Thurner, ``Multirelational organization of
	large-scale social networks in an online world,'' \emph{Proceedings of the
		National Academy of Sciences}, vol. 107, no.~31, pp. 13\,636--13\,641, 2010.
	
	\bibitem{chowdhury2020effect}
	S.~N. Chowdhury, D.~Ghosh, and C.~Hens, ``Effect of repulsive links on
	frustration in attractively coupled networks,'' \emph{Physical Review E},
	vol. 101, no.~2, p. 022310, 2020.
	
	\bibitem{gade2013frustration}
	P.~M. Gade and G.~Rangarajan, ``Frustration induced oscillator death on
	networks,'' \emph{Chaos: An Interdisciplinary Journal of Nonlinear Science},
	vol.~23, no.~3, p. 033104, 2013.
	
	\bibitem{lucarini2016extremes}
	V.~Lucarini, D.~Faranda, J.~M.~M. de~Freitas, M.~Holland, T.~Kuna, M.~Nicol,
	M.~Todd, S.~Vaienti \emph{et~al.}, \emph{Extremes and recurrence in dynamical
		systems}.\hskip 1em plus 0.5em minus 0.4em\relax John Wiley \& Sons, 2016.
	
	\bibitem{ray2019intermittent}
	A.~Ray, S.~Rakshit, D.~Ghosh, and S.~K. Dana, ``Intermittent large deviation of
	chaotic trajectory in Ikeda map: Signature of extreme events,'' \emph{Chaos:
		An Interdisciplinary Journal of Nonlinear Science}, vol.~29, no.~4, p.
	043131, 2019.
	
	\bibitem{ansmann2013extreme}
	G.~Ansmann, R.~Karnatak, K.~Lehnertz, and U.~Feudel, ``Extreme events in
	excitable systems and mechanisms of their generation,'' \emph{Physical Review
		E}, vol.~88, no.~5, p. 052911, 2013.
	
	\bibitem{ray2020extreme}
	A.~Ray, A.~Mishra, D.~Ghosh, T.~Kapitaniak, S.~K. Dana, and C.~Hens, ``Extreme
	events in a network of heterogeneous Josephson junctions,'' \emph{Physical
		Review E}, vol. 101, no.~3, p. 032209, 2020.
	
	\bibitem{ray2020understanding}
	A.~Ray, S.~Rakshit, G.~K. Basak, S.~K. Dana, and D.~Ghosh, ``Understanding the
	origin of extreme events in El Ni{\~n}o southern oscillation,''
	\emph{Physical Review E}, vol. 101, no.~6, p. 062210, 2020.
	
	\bibitem{cousins2015unsteady}
	W.~Cousins and T.~P. Sapsis, ``Unsteady evolution of localized unidirectional
	deep-water wave groups,'' \emph{Physical Review E}, vol.~91, no.~6, p.
	063204, 2015.
	
	\bibitem{guth2019machine}
	S.~Guth and T.~P. Sapsis, ``Machine learning predictors of extreme events
	occurring in complex dynamical systems,'' \emph{Entropy}, vol.~21, no.~10, p.
	925, 2019.
	
	\bibitem{dematteis2018rogue}
	G.~Dematteis, T.~Grafke, and E.~Vanden-Eijnden, ``Rogue waves and large
	deviations in deep sea,'' \emph{Proceedings of the National Academy of
		Sciences}, vol. 115, no.~5, pp. 855--860, 2018.
	
	\bibitem{grooms2014stochastic}
	I.~G. Grooms and A.~J. Majda, ``Stochastic superparameterization in a
	one-dimensional model for wave turbulence,'' \emph{Communications in
		Mathematical Sciences}, vol.~12, no.~3, pp. 509--525, 2014.
	
	\bibitem{majda2014conceptual}
	A.~J. Majda and Y.~Lee, ``Conceptual dynamical models for turbulence,''
	\emph{Proceedings of the National Academy of Sciences}, vol. 111, no.~18, pp.
	6548--6553, 2014.
	
	\bibitem{cai2001dispersive}
	D.~Cai, A.~J. Majda, D.~W. McLaughlin, and E.~G. Tabak, ``Dispersive wave
	turbulence in one dimension,'' \emph{Physica D: Nonlinear Phenomena}, vol.
	152, pp. 551--572, 2001.
	
	\bibitem{majda2015intermittency}
	A.~J. Majda and X.~T. Tong, ``Intermittency in turbulent diffusion models with
	a mean gradient,'' \emph{Nonlinearity}, vol.~28, no.~11, p. 4171, 2015.
	
	
	\bibitem{sornette2009dragon}
	D.~Sornette, ``Dragon-kings, Black Swans and the Prediction of Crises,''
	\emph{Swiss Finance Institute, Swiss Finance Institute Research Paper Series}, vol.~2, doi:10.2139/ssrn.1596032, 2009.
	
	\bibitem{feigenbaum2001statistical}
	J.~A. Feigenbaum, ``A statistical analysis of log-periodic precursors to
	financial crashes,'' \emph{Quantitative Finance}, vol.~1, pp. 346--360, 2001.
	
	\bibitem{bonatto2011deterministic}
	C.~Bonatto, M.~Feyereisen, S.~Barland, M.~Giudici, C.~Masoller, J.~R.~R. Leite,
	and J.~R. Tredicce, ``Deterministic optical rogue waves,'' \emph{Physical
		Review Letters}, vol. 107, no.~5, p. 053901, 2011.
	
	\bibitem{dysthe2008oceanic}
	K.~Dysthe, H.~E. Krogstad, and P.~M{\"u}ller, ``Oceanic rogue waves,''
	\emph{Annu. Rev. Fluid Mech.}, vol.~40, pp. 287--310, 2008.
	
	\bibitem{slunyaev2009rogue}
	A.~Slunyaev, C.~Kharif, and E.~Pelinovsky, \emph{Rogue Waves in the
		Ocean}.\hskip 1em plus 0.5em minus 0.4em\relax Springer, Berlin, 2009.
	
	\bibitem{chowdhury2019synchronization}
	S.~N. Chowdhury, S.~Majhi, M.~Ozer, D.~Ghosh, and M.~Perc, ``Synchronization to
	extreme events in moving agents,'' \emph{New J. Phys.}, vol.~21, no.~7, p.
	073048, 2019.
	
	\bibitem{buscarino2006dynamical}
	A.~Buscarino, L.~Fortuna, M.~Frasca, and A.~Rizzo, ``Dynamical network
	interactions in distributed control of robots,'' \emph{Chaos: An
		Interdisciplinary Journal of Nonlinear Science}, vol.~16, no.~1, p. 015116,
	2006.
	
	\bibitem{pecora2014cluster}
	L.~M. Pecora, F.~Sorrentino, A.~M. Hagerstrom, T.~E. Murphy, and R.~Roy,
	``Cluster synchronization and isolated desynchronization in complex networks
	with symmetries,'' \emph{Nature communications}, vol.~5, no.~1, pp. 1--8,
	2014.
	
	\bibitem{stilwell2006synchronization}
	D.~J. Stilwell, E.~M. Bollt, and D.~G. Roberson, ``Synchronization of
	time-varying networks under fast switching,'' in \emph{2006 American Control
		Conference}.\hskip 1em plus 0.5em minus 0.4em\relax IEEE, 2006, pp. 6--pp.
	
	\bibitem{Nag_Chowdhury_2020}
	S.~N. Chowdhury and D.~Ghosh, ``Hidden attractors: A new chaotic system without
	equilibria,'' \emph{The European Physical Journal Special Topics}, vol. 229,
	no. 6-7, pp. 1299--1308, 2020.
	
	\bibitem{pecora1990synchronization}
	L.~M. Pecora and T.~L. Carroll, ``Synchronization in chaotic systems,''
	\emph{Physical Review Letters}, vol.~64, no.~8, p. 821, 1990.
	
	\bibitem{camilli2006bacterial}
	A.~Camilli and B.~L. Bassler, ``Bacterial small-molecule signaling pathways,''
	\emph{Science}, vol. 311, no. 5764, pp. 1113--1116, 2006.
	
	\bibitem{taylor2009dynamical}
	A.~F. Taylor, M.~R. Tinsley, F.~Wang, Z.~Huang, and K.~Showalter, ``Dynamical
	quorum sensing and synchronization in large populations of chemical
	oscillators,'' \emph{Science}, vol. 323, no. 5914, pp. 614--617, 2009.
	
	\bibitem{schroder2015transient}
	M.~Schr{\"o}der, M.~Mannattil, D.~Dutta, S.~Chakraborty, and M.~Timme,
	``Transient uncoupling induces synchronization,'' \emph{Physical Review
		Letters}, vol. 115, no.~5, p. 054101, 2015.
	
	\bibitem{tandon2016synchronizing}
	A.~Tandon, M.~Schr{\"o}der, M.~Mannattil, M.~Timme, and S.~Chakraborty,
	``Synchronizing noisy nonidentical oscillators by transient uncoupling,''
	\emph{Chaos: An Interdisciplinary Journal of Nonlinear Science}, vol.~26,
	no.~9, p. 094817, 2016.
	
	\bibitem{sorrentino2016complete}
	F.~Sorrentino, L.~M. Pecora, A.~M. Hagerstrom, T.~E. Murphy, and R.~Roy,
	``Complete characterization of the stability of cluster synchronization in
	complex dynamical networks,'' \emph{Science advances}, vol.~2, no.~4, p.
	e1501737, 2016.
	
	\bibitem{dixit2020static}
	S.~Dixit and M.~D. Shrimali, ``Static and dynamic attractive--repulsive
	interactions in two coupled nonlinear oscillators,'' \emph{Chaos: An
		Interdisciplinary Journal of Nonlinear Science}, vol.~30, no.~3, p. 033114,
	2020.
	
	\bibitem{cantelli1929sui}
	F.~P. Cantelli, ``Sui confini della probabilita,'' in \emph{Atti del Congresso
		Internazionale dei Matematici: Bologna del 3 al 10 de settembre di 1928},
	1929, pp. 47--60.
	
	\bibitem{ghosh2002probability}
	B.~Ghosh, ``Probability inequalities related to markov's theorem,'' \emph{The
		American Statistician}, vol.~56, no.~3, pp. 186--190, 2002.
	
	\bibitem{heagy1995desynchronization}
	J.~Heagy, T.~Carroll, and L.~Pecora, ``Desynchronization by periodic orbits,''
	\emph{Physical Review E}, vol.~52, no.~2, p. R1253, 1995.
	
	\bibitem{platt1993off}
	N.~Platt, E.~Spiegel, and C.~Tresser, ``On-off intermittency: A mechanism for
	bursting,'' \emph{Physical Review Letters}, vol.~70, no.~3, p. 279, 1993.
	
	\bibitem{cavalcante2013predictability}
	H.~L. d.~S. Cavalcante, M.~Ori{\'a}, D.~Sornette, E.~Ott, and D.~J. Gauthier,
	``Predictability and suppression of extreme events in a chaotic system,''
	\emph{Physical Review Letters}, vol. 111, no.~19, p. 198701, 2013.
	
	\bibitem{pisarchik2011rogue}
	A.~N. Pisarchik, R.~Jaimes-Re{\'a}tegui, R.~Sevilla-Escoboza,
	G.~Huerta-Cuellar, and M.~Taki, ``Rogue waves in a multistable system,''
	\emph{Physical Review Letters}, vol. 107, no.~27, p. 274101, 2011.
	
	\bibitem{akhmediev2010editorial}
	N.~Akhmediev and E.~Pelinovsky, ``Editorial--introductory remarks on
	“discussion \& debate: Rogue waves--towards a unifying concept?”,''
	\emph{The European Physical Journal Special Topics}, vol. 185, no.~1, pp.
	1--4, 2010.
	
	\bibitem{kac1947notion}
	M.~Kac, ``On the notion of recurrence in discrete stochastic processes,''
	\emph{Bulletin of the American Mathematical Society}, vol.~53, no.~10, pp.
	1002--1010, 1947.
	
	\bibitem{altmann2005recurrence}
	E.~G. Altmann and H.~Kantz, ``Recurrence time analysis, long-term correlations,
	and extreme events,'' \emph{Physical Review E}, vol.~71, no.~5, p. 056106,
	2005.
	
	\bibitem{eichner2007statistics}
	J.~F. Eichner, J.~W. Kantelhardt, A.~Bunde, and S.~Havlin, ``Statistics of
	return intervals in long-term correlated records,'' \emph{Physical Review E},
	vol.~75, no.~1, p. 011128, 2007.
	
	\bibitem{chen2018social}
	X.~Chen and F.~Fu, ``Social learning of prescribing behavior can promote
	population optimum of antibiotic use,'' \emph{Frontiers in Physics}, vol.~6,
	p. 139, 2018.
	
	\bibitem{chen2019imperfect}
	X.~C. and F.~Fu, ``Imperfect vaccine and hysteresis,'' \emph{Proceedings of the
		Royal Society B}, vol. 286, no. 1894, p. 20182406, 2019.
	
	\bibitem{rossler1976equation}
	O.~E. R{\"o}ssler, ``An equation for continuous chaos,'' \emph{Physics Letters
		A}, vol.~57, no.~5, pp. 397--398, 1976.
	
	\bibitem{yan2019design}
	B.~Yan, S.~He, and K.~Sun, ``Design of a network permutation entropy and its
	applications for chaotic time series and eeg signals,'' \emph{Entropy},
	vol.~21, no.~9, p. 849, 2019.
	
	\bibitem{he2020fractional}
	S.~He, K.~Sun, and X.~Wu, ``Fractional symbolic network entropy analysis for
	the fractional-order chaotic systems,'' \emph{Physica Scripta}, vol.~95,
	no.~3, p. 035220, 2020.
	
	\bibitem{yan2020multistability}
	B.~Yan, S.~He, and S.~Wang, ``Multistability and formation of spiral waves in a
	fractional-order memristor-based hyperchaotic l{\"u} system with no
	equilibrium points,'' \emph{Mathematical Problems in Engineering}, vol. 2020,
	2020.
	
\end{thebibliography}

\ifCLASSOPTIONcaptionsoff
  \newpage
\fi



\bibliographystyle{IEEEtran}
\end{document}